\documentclass[copyright,creativecommons]{eptcs}
\providecommand{\event}{MBT 2012}

\title{Model-Based Testing of Safety Critical Real-Time Control Logic Software}
\author{Yevgeny Gerlits \qquad\qquad Alexey Khoroshilov
\institute{Institute for System Programming of the Russian Academy of Sciences \\ Alexander Solzhenitsyn st. 25, 1009004 \\ Moscow, Russia}
\email{gerlits@ispras.ru \qquad\qquad khoroshilov@ispras.ru}
}
\def\titlerunning{Model-Based Testing of Safety Critical Real-Time Control Logic Software}
\def\authorrunning{Y. Gerlits \& A. Khoroshilov}

\usepackage{graphicx}
\usepackage{listings}
\begin{document}
\maketitle

\begin{abstract}
The paper presents the experience of the authors in model based testing of safety critical real-time control logic software. It describes specifics of the corresponding industrial settings and discusses technical details of usage of UniTESK model based testing technology in these settings. Finally, we discuss possible future directions of safety critical software development processes and a place of model based testing techniques in it.
\end{abstract}

\section{Introduction}
\label{s1}

Role of safety critical systems in our life is increasing at a rapid pace. The distinctive characteristic of such systems is that their failure can be very dangerous for people or environment. As a result, development of safety critical systems is usually regulated by government agencies according to appropriate safety certification standards (e.g. DO-178B for avionics, BS EN 50128 for railways, IEC 60880 for nuclear, IEC 61508 for industry, IEC 62304 for medical devices). Requirements of these specifications make development and verification of safety critical software noticeably different from traditional processes.

The key differences that had an influence on our experience of model based testing development are as follows:
\begin{itemize} \itemsep1pt \parskip0pt \parsep0pt
\item requirement and architecture documents are carefully developed and maintained up to date;
\item there is a need for requirements based testing including explicit requirements traceability;
\item there is a need for source code coverage measurements depending on criticality level of a component under test;
\item tool qualification is required including tools used to automate verification processes.
\end{itemize}

Other important elements of the standards such as safety analysis affect our work to a very little degree.

Usually requirement documents consist of two levels: HLR (high level requirements) and LLR (low level requirements). HLR describe what is expected behavior of a component. LLR and software design documents provide sufficiently detailed description how to implement HLR. LLR based tests covers source code well even in terms of advanced source code coverage metrics such as MC/DC \cite{1} because source code is usually very close to LLR. Compliance between LLR and HLR is mostly verified using manual formal inspections. One more HLR verification technique is system-level tests based on HLR.

It is well known that model based testing works well when high quality testing is required. There are specialized model based testing technologies for real-time systems based on explicit automata definition, e.g. RT Tester \cite{2}, UPPAAL-TRON \cite{3}, Timed-TorX \cite{4} and TTG \cite{5}. On the other hand, there are model based functional testing technologies such as SpecExplorer \cite{6} and UniTESK \cite{8} that automates test sequence generation from implicitly defined automaton models. If advanced test coverage is needed and model automata are rather complex, these approaches are more efficient than manual test development. In comparison with explicit automata definition, the implicit one is less tend to lose clarity and accuracy when the complexity of the functionality under test increases because the increase in complexity results in the growth of the number of states and transitions in the automaton model of the SUT (system under test). Information about usage of these approaches in the domain of real time safety critical software is very limited, so we present our experience regarding usage of UniTESK model based testing technology to real time systems with rather complex functionality.

The rest of the paper is organized as follows. Section~\ref{s2} gives a short informal description of real time logic control subsystems. Section~\ref{s3} briefly describes the key ideas of the UniTESK model based testing technology. Section~\ref{s4} provides technical details of our usage of UniTESK technology for testing of safety critical real time logic control subsystems. Finally section~\ref{s5} outlines future works and possible improvements in development processes of safety critical software.

\section{Real-time logic control subsystems}
\label{s2}

In this section we give a short informal description of the object under test and its environment, i.e. the target real-time logic control subsystem  and the whole architecture of the RTES (real time embedded system) software respectively. We also provide a simple example of an iron automatic shut-off control subsystem. This subsystem is used in the subsequent sections to illustrate the most complicated aspects of our testing approach for control subsystems of RTES.

The RTES software consists of a number of subsystems. The responsibility of control subsystems includes analysis of the input data, decision making on the RTES reaction and generation of the output data needed to perform the reaction. Other subsystems are not limited but usually include subsystems that produce the input data for control subsystems and subsystems that process the output data produced by control subsystems. As a rule, subsystems producing the input data for control subsystems usually process the output data of some sensors and subsystems processing the output data produced by control subsystems usually perform the reaction of the RTES.

The subsystems being part of the RTES software are supposed to be called in the global control loop, i.e. they are subsequently called on each turn of the loop. The next turn of the control loop does not begin after the last subsystem has finished its work, but it starts after a certain period of time has expired. Let us refer to a turn of the control loop as a \emph{cycle} and to the period of time after which the next cycle begins as the \emph{cycle period}. The total execution time of all the subsystems on each cycle may not exceed the established cycle period for the RTES, otherwise the behavior of the RTES software is considered to be incorrect.

\begin{figure}
  \caption{Flow-chart of iron automatic shut-off control subsystem}
  \label{flow_chart}
  \begin{center}
  	\includegraphics[height=120px]{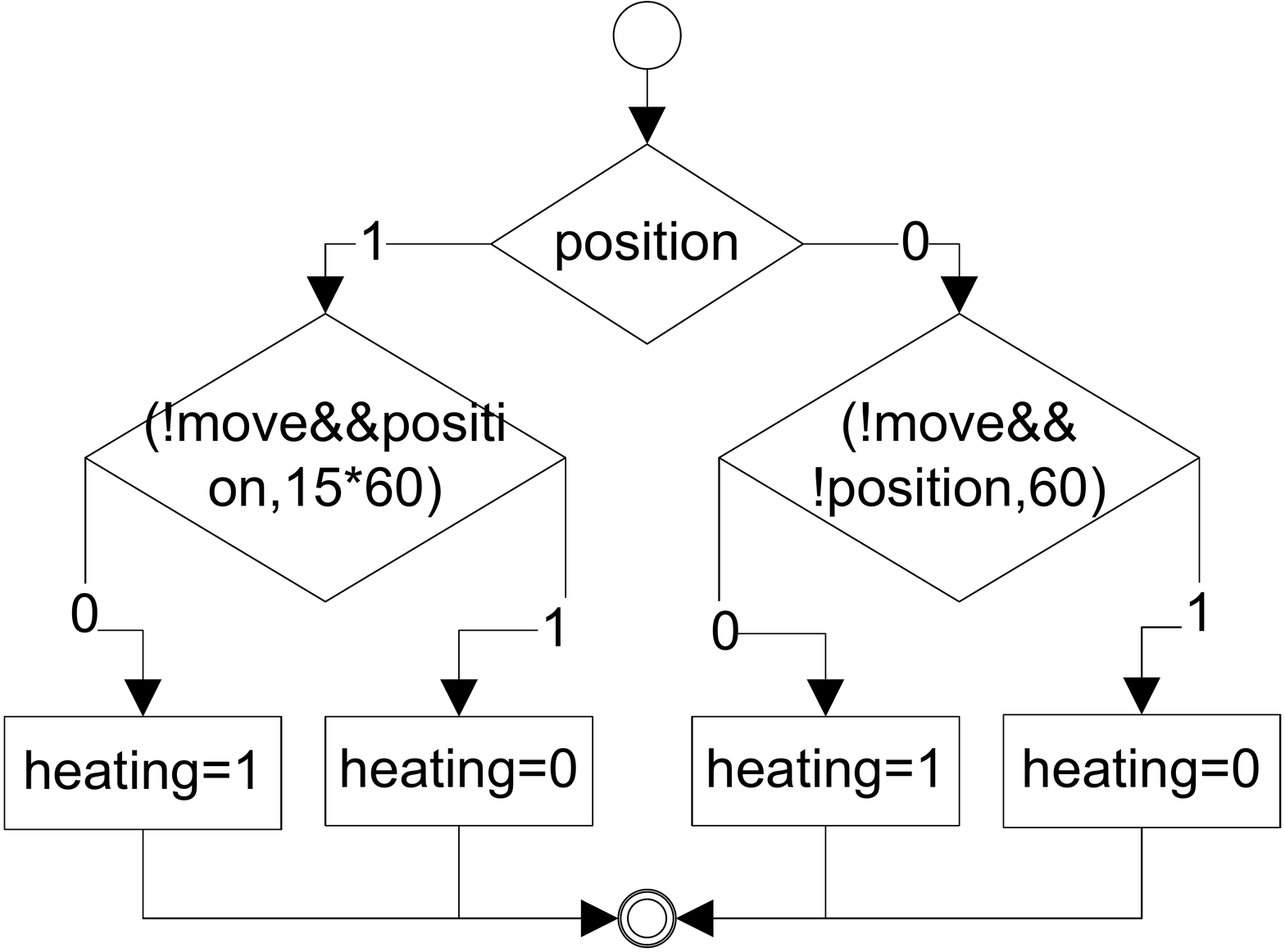}
  \end{center}
\end{figure}

Control subsystems of RTES consist of a number of decision making algorithms. The base part of a decision making algorithm consists of a scheme of branch instructions which can often be very complex. Figure~\ref{flow_chart} contains a simple iron automatic shut-off control subsystem. The names move and position are the input parameters of the subsystem. The parameter $move$ takes the value of 1 if the iron movement sensor detected that the iron has been moved since the beginning of the current cycle; otherwise the parameter takes the value of 0. The parameter $position$ takes the value of 1 if the iron position sensor detected that the iron is in the vertical position on the current cycle; otherwise the parameter takes the value of 0. The name $heating$ is the only output parameter of the subsystem. The value of 0 of this parameter prevents the iron sole from being heated; the value of 1 allows it to be heated.

Control subsystems of RTES use internal state variables to save data between cycles. Some state variables may be accessible for reading to other subsystems, some may not. The iron automatic shut-off control subsystem is simple. It does not have any state variables.

The key feature in designing of decision making logic in control subsystems of RTES is the possibility to build conditions in branch instructions on the basis of the notion of time. An example of a temporal condition is a Boolean formula which is required to be true for a period of time. Two such temporal conditions are used in the flow chart in figure~\ref{flow_chart}. The condition $(!move \&\& !position, 60)$ evaluates to true if the Boolean formula $(!move \&\& !position)$ has evaluated to true for 60 seconds.

How does a control subsystem of RTES calculate that a Boolean formula keeps the value of true for a period of time? Let \emph{system time} be the period of time passed since the start of the RTES execution. The value of the system time is fixed by the kernel of the RTES in the beginning of each cycle and remains fixed for all the subsystems of the RTES during the cycle period. The system time being constant during the cycle period is a natural architectural decision usually taken in the software design stage as it introduces determinism into the behavior of the RTES software. To calculate that a Boolean formula keeps the value of true for a period of time, the control subsystem evaluates the value of the Boolean formula on each cycle and using the system time accumulates the time interval during which the Boolean formula keeps the value of true from cycle to cycle.

\section{UniTESK technology}
\label{s3}

In this section we briefly describe the UniTESK testing approach for software systems which provide a synchronous interface \cite{8}. An interface is considered to be synchronous if the next \emph{stimulus} may only be applied after the \emph{reaction} to the previous one has been received. The notions of the stimulus and the reaction are parts of the UniTESK terminology. Real life examples of a stimulus applied to the SUT are a method call or a form submission. Examples of a reaction of the SUT include the return value of a method call or a web page reload.

The SUT is considered as a black box in UniTESK and is supposed to provide a number of \emph{interface functions} to access its functionality. The SUT may have an internal state. A call to an interface function with a set of values of the parameters is considered as an application of a stimulus to the SUT. It can result in a change in the internal state of the SUT and if the called interface function specifies the return value, it is returned and is considered as a reaction.

\begin{figure}
  \caption{UniTESK test system architecture}
  \label{unitesk_test_suite_arch}
  \begin{center}
  	\includegraphics[height=100px]{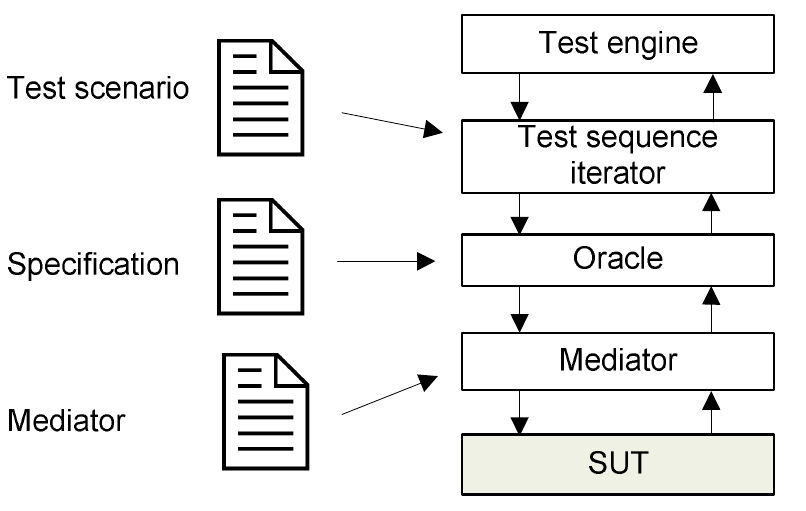}
  \end{center}
\end{figure}

Figure~\ref{unitesk_test_suite_arch} contains the universal UniTESK test system architecture. \emph{Test engine} is a library component which implements a traversal algorithm for the abstract automaton. Information which can only be supplied by the tester is concentrated in three components: \emph{oracle}, \emph{mediator} and \emph{test sequence iterator}. Compact formal descriptions are proposed for these components: \emph{specification}, \emph{mediator} and \emph{test scenario} accordingly. The formal descriptions are developed in programming languages which extend industrial programming languages with a number of syntactic constructions. Those are the \emph{specification extensions} of C and Java programming languages.\footnote{The renunciation of specification extensions in favor of natural language constructs becomes a new trend in the evolution of UniTESK. The number of supported languages is growing. C++ and Python implementations of UniTESK are now available.}

Functional requirements are formalized in specifications from which oracles are generated.  Oracles check the reactions of the SUT to the applied stimuli. The requirements specification technique used in UniTESK is based on the well-known Design by contract method \cite{9}. The state of the specification models the state of the SUT. \emph{State invariants} represent requirements to which all the specification states obtained during the execution of the test system must satisfy. The specification declares a \emph{specification function} for each interface function being tested. The parameters of a specification function model the parameters of the corresponding interface function. The requirements to the behavior of the interface function are formalized in the corresponding specification function in the form of a \emph{precondition} and \emph{postcondition} of the interface function call. The specification also formulates a coverage criterion on the basis of the structure of the formalized requirements \cite{10}.

The precondition describes legal calls to the interface function on the basis of the current state of the specification and the parameter values of the specification function. A \emph{specification stimulus} in UniTESK is a call to a specification function. It results in the check of the precondition. Violation of the precondition indicates that the test system tried to perform an illegal call to the interface function. Before checking the postcondition, the test system saves the current specification state and calls the mediator function with the set of values of the specification function parameters.

The formal description called mediator is intended to bind a specification to the SUT and must contain a \emph{mediator function} for each specification function in specification. The signature and the type of the return value of the mediator function must be identical to those of the corresponding specification function. The mediator function transforms the specification representation of the input parameters of the interface function into the representation of the SUT, calls the interface function, transforms the return value of the interface function call into the representation of the specification, synchronizes the specification state with the state of the SUT and returns the specification representation of the return value. The specification representation of the return value is considered in the postcondition of the specification function as the return value of the specification function and is called \emph{specification reaction}.

After the mediator function call has completed, the test system initiates the check of the postcondition of the specification function. The specification reaction and the specification state after the call to the interface function are verified in the postcondition on the basis of the values of the input parameters of the specification function and so called the pre-state of the specification, i.e. the specification state at the time of the specification function call, which we know was preliminary saved.

The test scenario formalizes a model of the SUT in the form of the abstract automaton. The abstract automaton is used to automatically generate a sequence of \emph{test actions}. The test scenario describes the abstract automaton implicitly. The base part of the test scenario consists of the following functions: a number of \emph{scenario functions}, the \emph{initialization function}, the \emph{state generation function}, and the \emph{finalization function}. A scenario function is responsible for organization of calls to specification functions. A simple case implies the development of one scenario function for each specification function. Although one scenario function may perform a number of calls to specification functions, one scenario function is considered to describe one \emph{test action} from the point of view of the test scenario, i.e. a call to a scenario function is considered to perform one test action.\footnote{If a scenario function contains some iteration statements, it generally describes a number of test actions for each state of the abstract automaton.}

The abstract automaton of the SUT is being dynamically constructed during the execution of the test scenario. The initialization function moves the SUT and its specification into their initial states which must certainly be synchronized. After the initialization function has finished its execution, the state generation function is called. This function generates a state of the abstract automaton on the basis of the current state of the specification and/or the SUT. The test engine component either performs a test action which is defined for the current state of the abstract automaton and has not been performed yet or it performs an action defined for the current state to get to a state for which a test action is defined and has not been performed yet. Let the test engine applies a test action which has not been performed yet. A new transition is created from the state where the test action has been performed. The transition is marked with the test action performed.\footnote{If the scenario function being called does not contain any iteration statements, the transition is marked with the name of the scenario function; otherwise the transition is marked with the name of the scenario function complemented with the set of values of all the iteration variables.} The state of the SUT might change because of the test action. The state generation function is called to generate the end state for the transition. Construction of the test sequence terminates when the test engine has performed all the test actions defined for all the states reached during the execution of the test scenario. The finalization function is called at the end of the testing process. This function performs any actions related to the end of the testing process. For instance, it can release the allocated resources.

The main goal of the test scenario developer is to specify a set of states of the abstract automaton and a number of test actions for each state so that if the test system performs all the test actions, the target coverage criteria will be satisfied. The test scenario developer has to take into account some restrictions that the test engine component imposes on the abstract automaton. Although the abstract automaton is being dynamically constructed during the execution of the test scenario, the restrictions are imposed on the final abstract automaton. There are two main implementations of the test engine: a test engine that can traverse deterministic abstract automata \cite{11} and a test engine that can traverse nondeterministic abstract automata \cite{12}. The test engine for deterministic abstract automata demands the final abstract automaton to be finite, deterministic and strongly connected. The test engine for nondeterministic abstract automata demands the final automaton to be finite and to contain a deterministic, strongly connected, total, spanning subautomaton.

\section{Application of UniTESK for real-time logic control subsystems}
\label{s4}

The original UniTESK test system architecture undergoes some changes when applied to control subsystems of RTES. In this section we describe those changes and some features in the implementations of the individual architectural components.

\subsection{Test system architecture}

\begin{figure}
  \caption{Test system architecture for CSUT}
  \label{unitesk_for_csut}
  \begin{center}
  	\includegraphics[width=260px]{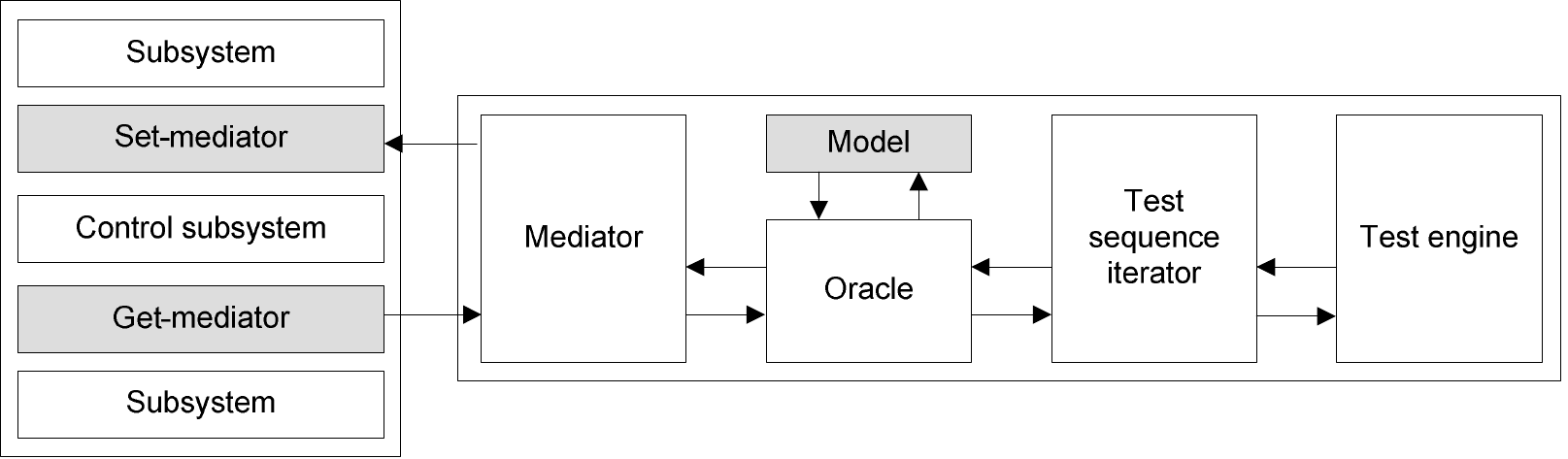}
  \end{center}
\end{figure}

Software and hardware resources are often limited in the computer system on which the software of RTES must run. Running resource consuming test system components on the target computer system is impossible therefore the test system should be implemented as a separate program in order that it may be run on a separate computer system. The architecture of the test system for functional testing of control subsystems of RTES is represented in figure~\ref{unitesk_for_csut}. It is composed of two programs which communicate synchronously. The software of RTES with two injected subsystems named set-mediator and get-mediator is in the figure~\ref{unitesk_for_csut} on the left. The main part of the test system is on the right. We will further refer to it simply as the test system.

The whole composition operates in the following way. At the beginning of each cycle, the test system generates a set of values for the input parameters of the CSUT (control subsystem under test). The mediator component sends the generated set of parameter values to the set-mediator subsystem which initializes the input parameters of the CSUT with them. After the CSUT has finished its execution, the get-mediator subsystem reads the values of the output parameters and accessible state variables of the CSUT, gets the system time at the current cycle and transfers all the data to the mediator component. The test system verifies the behavior of the CSUT at the current cycle and the whole composition goes on to the next cycle.

Requirements are formalized in UniTESK in the form of a precondition and a postcondition as implicit specification. Meanwhile, LLR usually transform inputs to outputs explicitly. Development of an explicit behavioral model from LLR is easier than development of an implicit model therefore a new component named model  is introduced into the test system architecture which implements the behavioral model of the CSUT explicitly. More precisely, the model is intended to produce reference values for the input parameters and the state variables of the CSUT at each cycle on the basis of the state and the input parameters of the CSUT in the model representation. The postcondition uses the referenced values produced by the model to compare them with the values produced by the CSUT.

\subsection{Model}

The model is intended to represent LLR in a formal way. Such a representation usually preserves the structure of the branch instructions fixed in LLR. Each decision control algorithm from LLR is mapped to a separate function. One function is designed to be the entry point to the model. Let us call it \emph{interface function of the model}. A model representation is developed for the input and output parameters and state variables of the CSUT. The input parameters in the model representation become part of the input parameters of the interface function of the model. The output parameters in the model representation become part of the output parameters of the interface function. The state variables in the model representation become part of both the input and output parameters of the interface function.

The input parameters and state variables of the CSUT in the model representation are not enough to evaluate temporal conditions in the model therefore the arguments of the interface function of the model must be complemented with some new parameters. By way of example, let us design a model representation for the temporal conditions used in the iron automatic shut-off control subsystem and complement the arguments of the interface function with parameters that would be enough to evaluate the modeled conditions. Here are the temporal conditions in the flow-chart of the iron automatic shut-off control subsystem:
\begin{enumerate}\itemsep1pt \parskip0pt \parsep0pt
\item $ (!move \&\& !position, 60) = (!move, 60) \&\& (!position, 60); $
\item $ (!move \&\&  position, 15*60) = (!move, 15*60) \&\& (position, 15*60). $
\end{enumerate}

The list of different temporal predicates assigned to a unique identifier where each predicate is a member of a temporal condition:
\begin{enumerate}\itemsep1pt \parskip0pt \parsep0pt
\item $ move\_eq\_f\_t1 = (!move, 60); $
\item $ position\_eq\_f\_t1  = (!position,  60); $
\item $ move\_eq\_f\_t2 = (!move, 15*60); $
\item $ position\_eq\_t\_t2  = (position, 15*60). $
\end{enumerate}

Listing~\ref{ids_of_temporal_predicates} represents a structure type which combines the identifiers of the temporal predicates. The arguments of the interface function of the model must be complemented with a parameter of this type. Let us call this parameter $time\_flags$.

\lstset{caption=Structure type combining identifiers of temporal predicates, label=ids_of_temporal_predicates}
\begin{lstlisting}
typedef struct{
  bool move_eq_f_t1;
  bool move_eq_f_t2;
  bool position_eq_f_t1;
  bool position_eq_t_t2;
} t_time_flags;
\end{lstlisting}

Listing~\ref{model} represents the source code of the model for the iron automatic shut-off control subsystem. The two temporal conditions are formalized on the basis of the identifiers of the temporal predicates. The value for the parameter $time\_flags$ is calculated by the mediator component on the basis of the previous interactions with the CSUT. This question is discussed later.

\lstset{caption=Model for iron automatic shut-off control subsystem, label=model}
\begin{lstlisting}
bool model(bool position, t_time_flags time_flags){
  if(position)
    if(time_flags.move_eq_f_t2 && time_flags.position_eq_t_t2)	    
      return false;
    else return true;
  else
    if(time_flags.move_eq_f_t1 && time_flags.position_eq_f_t1)	    
      return false;
    else return true;
}
\end{lstlisting}

\subsection{Specification}

The specification contains only one specification function because control subsystems of RTES are considered to provide one interface function. The input and output parameters of the CSUT in the model representation become the input and output parameters of the specification function accordingly. The state of the specification includes the state variables of the CSUT in the model representation. If temporal conditions are used in the branch instructions of the CSUT, the state of the specification is complemented with variables of the data types designed to evaluate the temporal conditions in the model.

Let us consider the postcondition of the specification function. At first, the interface function of the model is called to get referenced values for the output parameters and state variables of the CSUT in the model representation. Those parameters of the interface function of the model which correspond to the state variables of the CSUT get the pre-values of their counterparts in the specification state. Those parameters which correspond to temporal conditions get the post-values of their counterparts in the specification state. The verdict in the postcondition is returned on the basis of the compare of the output parameters and the state variables of the CSUT in the model representation with their referenced counterparts calculated by the model. Listing~\ref{specification} contains the specification for the iron automatic shut-off control subsystem.

\lstset{caption=Specification for iron automatic shut-off control subsystem, label=specification}
\begin{lstlisting}
t_time_flags time_flags;
specification bool spec(bool move, bool position) writes time_flags{
    post {return spec == model(position, time_flags);}
}
\end{lstlisting}

Some part of HLR can be represented in UniTESK tests as data invariants and extra checks in postconditions. It allows providing verification of HLR using the same LLR-based module-level tests.

UniTESK provides a number of coverage criteria on the basis of the structure of the requirements formalized in specifications \cite{10}. When UniTESK is applied for control subsystems of RTES, it is less labor intensive to measure the coverage of the source code of the model. One can easily use a structural coverage metric based on the control flow graph like the branch coverage, condition coverage, modified condition decision coverage and etc \cite{1, 13}.

\subsection{Mediators}

The previous section describes internals of the specification for control subsystems of RTES. In this section we consider the mediator component for this specification and refine the communication protocol between the mediators.

\subsubsection{Communication protocol between mediators}

The central part of the mediator is a single mediator function, the signature of which is determined by the signature of the specification function. The mediator function transforms the model representation of the input parameters of the CSUT into the representation of the CSUT, sends the values of the transformed parameters to the set-mediator subsystem and goes to the idle mode where it waits for the values of the output parameters and state variables of the CSUT and the system time at the current cycle. At each cycle the set-mediator subsystem starts and finishes its execution before the control subsystem has started. At first the set-mediator subsystem goes to the idle mode where it waits for values of the input parameters of the CSUT. After the values have been received, the set-mediator subsystem initializes the input parameters of the CSUT with them and finishes its execution. After the control subsystem has finished its execution, the get-mediator subsystem starts. At first it reads the values of the output parameters and accessible state variables of the CSUT. It also reads the value of the system time at the current cycle or calculates it. The get-mediator subsystem sends all the collected data to the mediator component and finishes its execution. The mediator function receives the data sent by the get-mediator subsystem. The values of the output parameters of the CSUT are transformed into the representation of the model. They will be returned as the return value of the mediator function at the end of its execution. Before doing this the mediator function starts to synchronize the specification state with the state of the CSUT.

\subsubsection{Synchronization of specification state with implementation state}

The state variables of the CSUT received from the get-mediator subsystem are transformed into the representation of the model. After that they are assigned to the specification state variables which model the state variables of the CSUT. Some state variables of the CSUT may not be accessible for reading to other subsystems including the get-mediator subsystem. The model representation for the inaccessible state variables is evaluated according to the principle of working with the hidden state, i.e. in the assumption that the CSUT operates without errors in compliance with the specification.

The mediator component should also calculate values for the specification state variables used to model temporal conditions. By way of example, listing~\ref{algorithm} contains an algorithm which evaluates the temporal predicate $(p == VAL, T)$. The mediator function is supposed to execute the algorithm at each cycle. The algorithm uses the following labels: $p$ is an input parameter of the CSUT, $VAL$ is a value of $p$, $T$ is a time interval during which p should preserve $VAL$, $sys\_time$ is the system time at the current cycle, $p\_sys\_time$ is the system time since which $p$ has preserved $VAL$ or -1 if $p$ is not equal to $VAL$.

\lstset{caption=Algorithm evaluates temporal predicate, label=algorithm}
\begin{lstlisting}
int p_sys_time = -1;
int get_p_eq_VAL_T(int p, int VAL, int T, int sys_time, int* p_sys_time){
  if(p != VAL) {*p_sys_time = -1;  return 0;}
  else{
    if(*p_sys_time == -1) *p_sys_time = sys_time;
    if(sys_time – *p_sys_time >= T) return 1;
    else return 0;
  }
}
\end{lstlisting}

\subsubsection{Control of cycle period}

The set-mediator subsystem goes to the idle mode at the first step of its execution. Being in the idle mode it is waiting for values for the input parameters of the CSUT. It is not ruled out that this delay becomes too long at a cycle so that the total time of all subsystems execution will exceed the cycle period. In this case the behavior of the RTES software is considered incorrect by reason which does not depend on the RTES software. Let us propose two solutions for the problem specified.

The first solution is the following. The cycle period can be increased so that the total time of all subsystems execution would not exceed the cycle period. In order to enhance the accuracy of the method, the test system might control that the total time does not exceed the cycle period at each cycle. The test system will indicate at the end of the testing process whether this condition has hold or not. If it has hold, the tester may start to analyze the test reports produced. If the condition has not hold, the cycle period must be increased.

The second solution requires that the RTES software implements two features: the \emph{streaming mode} and the system time accessible for writing to subsystems. The streaming mode makes the next cycle start immediately after the last subsystem has finished its execution. The streaming mode implies that the cycle period is not controlled and the subsystems may exceed it without any reaction of the RTES engine. The RTES software usually implements the streaming mode as it is required to easy debug the RTES software or it can be implemented exclusively to facilitate testing. If the system time is accessible for writing, one can implement a subsystem which starts at the beginning of each cycle and sets the system time at the current cycle to the sum of the system time at the previous cycle and the cycle period. From one hand, the two features can be used to simulate usual conditions of operation when each cycle starts right after the cycle period has expired. On the other hand, the total time of all subsystems execution cannot exceed the cycle period.

\subsection{Test scenario}

Development of test scenarios is seemed to be the most complicated task. Let us explain it in this section by an example, i.e. by developing principal parts of a test scenario for the iron automatic shut-off control subsystem. At first, the target coverage criterion should be established. Let it be the branch coverage of the model. The main goal of the test scenario developer consists in designing such an abstract automaton, traversal of which by the test engine satisfies the target coverage criterion.

\subsubsection{States of abstract automaton}

If the number of possible states of the specification is not high, i.e. several hundreds, the state generation function of the test scenario may take the current state of the specification as the state of the abstract automaton. If the number of possible states of the specification is dramatically high or even infinite, the approach of state generalization is used in UniTESK to construct the states of the abstract automaton. According to this approach the states of the specification are partitioned into equivalence classes which are used as the states of the abstract automaton. A universal method called coverage-targeted reduction of the model \cite{14} is used to design generalized states of the abstract automaton and test actions in scenario functions so that the traversal of the abstract automaton would satisfy the target coverage criterion. This method has already been applied in case of coverage criteria on the basis of the structure of the requirements in the specification. Here we give an example of its use in case of a structural criterion on the basis of the control flow graph, i.e. the target branch coverage of the model.

At the first step, a set of test cases is extracted. This set of test cases should satisfy the target coverage criterion. Using the flow-chart of the iron automatic shut-off control subsystem, one can find that the following test cases satisfy the branch coverage of the model:
\begin{enumerate}\itemsep1pt \parskip0pt \parsep0pt
\item $position \&\& (!move \&\& position, 15*60);$
\item $position \&\& !(!move \&\& position, 15*60);$
\item $!position \&\& (!move \&\& !position, 60);$
\item $!position \&\& !(!move \&\& !position, 60).$
\end{enumerate}

At the second step the test cases are considered as conditions which cut out subregions in the space of both possible states of the specification and possible values of the input parameters of the specification function:
\begin{enumerate}\itemsep1pt \parskip0pt \parsep0pt
\item $position \&\& move\_eq\_f\_t2 \&\& position\_eq\_t\_t2;$
\item $position \&\& !(move\_eq\_f\_t2 \&\& position\_eq\_t\_t2);$
\item $!position \&\& move\_eq\_f\_t1 \&\& position\_eq\_f\_t1;$
\item $!position \&\& !(move\_eq\_f\_t1 \&\& position\_eq\_f\_t1).$
\end{enumerate}

At the third step the subregions are projected on the space of possible states of the specification. Each projection extracts a subspace of states in which the corresponding test case is able to occur. In other words, a set of values for the input parameters may be found which covers the test case in each state from the projection. The conditions extracted at the previous step are projected on the following subspaces of states:
\begin{enumerate}\itemsep1pt \parskip0pt \parsep0pt
\item $move\_eq\_f\_t2 \&\& position\_eq\_t\_t2;$
\item $!(move\_eq\_f\_t2 \&\& position\_eq\_t\_t2);$
\item $move\_eq\_f\_t1 \&\& position\_eq\_f\_t1;$
\item $!(move\_eq\_f\_t1 \&\& position\_eq\_f\_t1).$
\end{enumerate}

\begin{figure}
  \caption{Generalized states of abstract automaton}
  \label{fsm_states}
  \begin{center}
  	\includegraphics[width=90px]{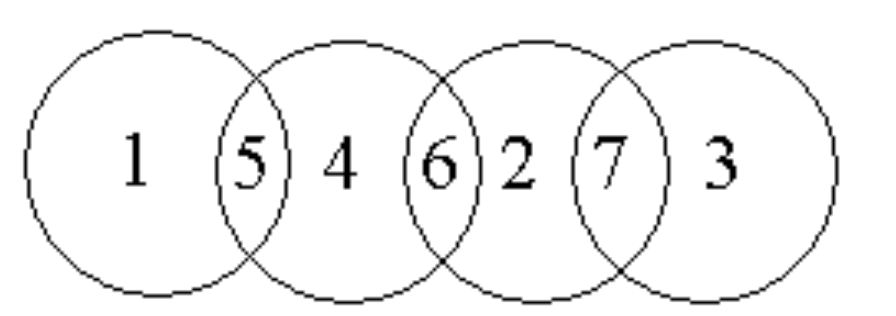}
  \end{center}
\end{figure}

The projections may have nonempty intersections. In other words, there are states in which several test cases may be covered depending on the values of the input parameters. At the fourth step all different intersections of the projections are taken. Figure~\ref{fsm_states} outlines this operation on the projections obtained at the previous step. The intersections are numbered from 5 to 7. The final set of generalized states partitions the whole space of states into equivalence classes because two calls to the specification function in different states from the same generalized state result in the same set of test cases being covered.

The state generation function must not explicitly take different intersections of the projections on the space of states. The current state of the specification and the set of conditions which specify the projections are enough to build the generalized state of the abstract automaton by returning the vector $(\alpha_1, \alpha_2, \ldots, \alpha_n) \in \{0,1\} : \alpha_i = 1$ if the specification state belongs to projection $i$; otherwise $\alpha_i = 0$.

Extraction of conditions which specify the projections of test cases to the state space of the specification may be a really complicated task. Long-term experience in development of test scenarios for software systems of different classes was investigated in the paper \cite{15}. An interesting idea consists in extracting states of the abstract automaton on the basis of the data structures which model the state of the SUT in the specification. The paper describes some approaches to development of test scenarios. Each approach is based on a widely used data structure. The approaches are represented in the form of design patterns. The statistics collected during that research conforms that the collected patterns are used in more than 80\% of the test scenarios investigated.

\subsubsection{Test actions}

Let the states of the abstract automaton have been developed with the method of coverage-targeted reduction of the model. The method of coverage-targeted reduction of the model describes the way in which the test actions in scenario functions must be defined. The test actions are considered as the equivalence classes of calls to the specification function by the target coverage criterion. In other words, one test action corresponds to a set of calls to the specification function where each call covers the same coverage item selected by the coverage criterion.

Any alternative method of defining test actions must result in the target coverage criterion satisfied. Having this in mind, we recommend developing scenario functions of two types. Scenario functions of type 1 make reachable states of the abstract automaton needed to satisfy the target coverage criterion. Scenario functions of type 2 iterate values of the parameters of the specification function in each state to maximize the value of the target coverage metric. This approach is effective in practice and is more intuitive.

\subsubsection{Restrictions imposed by test engine}

We have focused so far on the problem of designing such an abstract automaton, traversal of which would satisfy the target coverage criterion. It is not the only problem which usually appears during development of test scenarios. Restrictions imposed by the test engines on the final abstract automaton have to be in mind during development of both states and test actions. It is difficult to control the restrictions during development of the test scenario because the restrictions are imposed on the final abstract automaton but the abstract automaton of the SUT is being dynamically constructed during the testing process. The most complicated property to satisfy in practice is the determinacy of the abstract automaton and the determinacy of the spanning subautomaton. The method of coverage-targeted reduction of the model does not ensure that the final abstract automaton will satisfy one of these properties. The following methods can be used to modify the abstract automaton so that it would satisfy the determinacy or determinacy of the spanning subautomaton \cite{14, 15}: \emph{splitting of states, injection of connective transitions, generalization of transitions}. These methods do not directly concern control subsystems of RTES. We do not consider them in this paper.

\subsubsection{Test sequence reduction methods}

The total time of execution of test scenarios may be dramatically high in some cases. This happens because the number of test actions applied during execution of test scenarios is high. It should be noted here that one test action specified in a scenario function may be applied many times during execution of the test scenario because the traversal algorithm implemented in the test engine applies known test actions to get to a state where there are some test actions not applied yet. Decreasing the number of test actions is considered as a big problem because the target coverage criteria must be ultimately satisfied whatever is happened with the abstract automaton. 

The following methods are suggested to solve the problem: \emph{testing piecemeal, filtering of input parameters in scenario functions, enlargement of states}. These methods are based on a practical observation and at the same time a logical consideration according to which reducing the number of both possible states of the abstract automaton and test action specifications in scenario functions results in decreasing the number of test actions really applied during execution of the test scenario. We do not formalize the methods in this paper and do not formally proof their effectiveness, but describe them informally and argument why they are useful and usually deliver the expected result in practice.

\textbf{Testing piecemeal}. According to this method the CSUT is partitioned into logical parts. Each part is tested by a separate test scenario. Parts can be algorithms, flow-charts, parts of algorithms or even separate branch instructions. It depends on the complexity of the CSUT and the target coverage criteria. The test scenario for a part of the CSUT is expected not to define more states and test action specifications in scenario functions than the test scenario for the CSUT as a whole. Testing piecemeal results in the input parameters as well as the states of the CSUT gone over piecemeal therefore the total number of test actions performed by the test scenarios for the parts of the CSUT is expected not to exceed the length of the test sequence generated by the test scenario for the CSUT as a whole. If possible, the test scenarios for the parts of the CSUT can be run simultaneously to further reduce the total time of execution.

Implementation of the method implies that the test scenario takes control over the data and control flows in the CSUT. Some input parameters are set to values which ensure that the data and control flows reach the target part of the CSUT, the other parameters are gone over to satisfy the coverage of the target part of the CSUT.

Some coverage criteria do not allow for testing piecemeal because all structure of the SUT or requirements must be taken into account to satisfy the criteria. A typical example is MCC (Multiple Condition Coverage) \cite{1}.

\textbf{Filtering of input parameters in scenario functions}. Filters are conditions on a set of variables which filter out some sets of their values. The method reduces the number of test action specifications in scenario functions by applying filters in iteration statements to filter out those set of values for the input parameters of the CSUT which do not enlarge the coverage. The syntax rules for filters in iteration statements can be found in \cite{10}.

\textbf{Enlargement of states}. The goal of the method consists in reducing the number of states of the abstract automaton by re-generalizing the specification states so that the new set of generalized states would contain more specification states than the prevous set on the average. Let a test scenario is implemented for the iron automatic shut-off subsystem and the states of the specification act as the states of the abstract automaton. We can remember that the state of the specification is determined by the values of four temporal predicates therefore the upper bound for the number of possible states of the abstract automaton is 16. We obviously might calculate that only 9 states are really possible. As we showed the method of coverage-targeted reduction of the model results in 7 states of the abstract automation. Both 9 and 7 states cover the whole space of possible states of the specification. It is a true fact that we have constructed 7 states that are larger that 9 original states on the average.

\subsection{Practical results}

All solutions described in this paper were implemented and applied in projects on testing of avionics related control subsystems. As an example, we provide some details of our joint project with Russian System Corporation, where a system under test was the control subsystem of AAFSS (Airborne Active Flight Safety System) developed by Russian System Corporation.

AAFSS is designed to increase flight safety and effectiveness of the airborne complex. The system performs monitoring of the operational status of the airborne systems, survival facilities, operational conditions and adequacy of the behavior of the crew, as well as decision control on recovery of flight safety in critical situations.

The AAFSS software consists of a number of subsystems. The control subsystem is the most critical one because it takes all decisions in AAFSS. The control subsystem consists of about 10 decision control algorithms, 30 input parameters, 10 state variables, 10 output parameters and 80 temporal predicates. The functional requirements to the control subsystem are well-structured and carefully described mainly in flow-charts.

The numbers of the input parameters, state variables and temporal predicates do not give an alternative to the method of testing piecemeal. 15 test scenarios were developed for the control subsystem of AAFSS. Each scenario performs about 58000 test actions on the average. Some critical bugs were revealed by those test scenarios in development versions of control software.

We got added evidence that an important advantage of a model based testing technology complemented with a strong process of formalization of requirements consists in possibility to reveal bugs at earlier stages \cite{7}. In particular, some problems concerning correctness, unambiguity and completeness \cite{16, 17} of the decision control algorithms were revealed.

\section{Conclusions}
\label{s5}

The paper presents our experience in model based testing of control subsystems of safety critical RTES using the UniTESK testing technology. The input for testing process is LLR that describe implementation details of the control logic in an informal way. This description is usually similar to flow charts enriched by timing properties. The tests to be developed have to cover all LLR and have to provide appropriate source code coverage. The latter means that the tests have to cover all the decisions and/or conditions that basically inherited from LLR.

In these settings model based testing techniques demonstrate themselves as an efficient means to achieve the required coverage level in a semi-automated way. Basically, the UniTESK model based testing approach suggests to formalize LLR and then to take benefits from the automation of the verification activities on the base of this model. Formalization of  LLR allows to reveal issues in LLR themselves. Additionally, UniTESK allows to provide verification of HLR on the base of the same LLR-based module-level tests.

The next logical step could be to formalize LLR from the very beginning using some formal notation. The possible benefits of this step include:
\begin{itemize}\itemsep1pt \parskip0pt \parsep0pt
\item generation of source code from LLR;
\item generation of UniTESK models from LLR if testing against LLR is required;
\item earlier bug revealing in LLR during the formalization step;
\item partial automation of LLR verification against HLR.
\end{itemize}

There are already several tools available that support formalization of LLR. The most known of them are SCADE and Simulink. Qualified code generators have been developed for the models produced by these tools. If qualified code generators are available in a project, there is no need in LLR-based tests. In this case model based testing techniques can be valuable to automate test generation in the context of HLR verification. If qualified tools are not available in a project, LLR-based tests described above are still required for certification purposes. In this case most parts of the UniTESK test suite could be generated from LLR except for the test scenarios, where manual tuning is required to prevent the state explosion.

\bibliographystyle{eptcs}

\end{document}